\documentclass[twocolumn,showpacs,preprintnumbers,amsmath,amssymb,prb,superscriptaddress,aps,10pt]{revtex4-1}

\usepackage{graphicx}
\usepackage{float}
\usepackage{overpic}
\usepackage[usenames,dvipsnames]{color}
\usepackage{textcomp}

\usepackage{graphics}
\usepackage{graphicx}
\usepackage{epsfig,psfrag}

\newcommand{\comment}[1]{}

\begin{document}

\title{Kondo effect and the fate of bistability in molecular quantum dots 
with strong electron-phonon coupling}
\author{Juliane Klatt}
 \affiliation{Physikalisches Institut,
Albert-Ludwigs-Universit\"at Freiburg,\\
 Hermann-Herder-Stra{\ss}e 4, D-79104 Freiburg i. Br., Germany}
\author{Lothar M\"uhlbacher}
\affiliation{Physikalisches Institut,
Albert-Ludwigs-Universit\"at Freiburg,\\
 Hermann-Herder-Stra{\ss}e 4, D-79104 Freiburg i. Br., Germany}
  \affiliation{Freiburg Institute for Advanced Studies (FRIAS),
Albert-Ludwigs-Universit\"at Freiburg,\\
 Albertstra{\ss}e 19, D-79104 Freiburg i. Br., Germany}
\author{Andreas Komnik}
\affiliation{Institut f\"ur Theoretische Physik,
Ruprecht-Karls-Universit\"at Heidelberg,\\
 Philosophenweg 12, D-69120 Heidelberg, Germany}
 \affiliation{Physikalisches Institut,
Albert-Ludwigs-Universit\"at Freiburg,\\
 Hermann-Herder-Stra{\ss}e 4, D-79104 Freiburg i. Br., Germany}
 \affiliation{Freiburg Institute for Advanced Studies (FRIAS),
Albert-Ludwigs-Universit\"at Freiburg,\\
 Albertstra{\ss}e 19, D-79104 Freiburg i. Br., Germany}
 
\date{\today}

\begin{abstract}

We investigate the properties of the molecular quantum dot (Holstein-Anderson) model using numerical and analytical techniques. Path integral Monte Carlo simulations 
for the cumulants of the distribution function of the phonon coordinate reveal that at intermediate temperatures the effective potential for the oscillator exhibits two minima rather than a single one,
{\color{black} which can be understood as a signature of a bistability effect.}  A straightforward
adiabatic approximation turns out to adequately
  describe the properties of the system in
this regime. Upon lowering the temperature the two potential energy
  minima of the oscillator merge to a single one at the equilibrium position of
  the uncoupled system.
Using the parallels to the X-ray edge problem in metals 
we derive the oscillator partition function. 
It turns out to be identical to that of
the Kondo model, which is known to possess a universal low energy fixed point
characterized by a single parameter -- the Kondo temperature $T_K$.  We derive
{\color{black} an analogon of} $T_K$ for the molecular quantum dot model, present numerical evidence pointing
towards the appearance of the Kondo physics
and discuss experimental implications of the discovered phenomena.

\end{abstract}


\maketitle

In view of the recent progress in the field of microelectronic fabrication, which produces ever smaller electronic circuitry elements, it is reasonable to assume that the basic building blocks of the future nanoelectronics would be individual molecules.\cite{cuniberti}  Contrary to the solid-state based systems their internal degrees of freedom play a principal role.
 The most important ones are the vibrational degrees of freedom.\cite{GalperinRatnerNitzan,Huetzen} Although it is possible to model their effects with the help of a rather simple model -- the \emph{molecular quantum dot} (sometimes also referred to as \emph{Holstein-Anderson model}), its properties are still not understood in full detail [\onlinecite{cuevas2010molecular}]. One of the reasons is that the problem in general is not exactly solvable and many of the interesting regimes are not accessible analytically. In particular, some time ago it was predicted that when the electron-phonon coupling is sufficiently strong, such a molecular dot might possess a bistability regime. \cite{my,Alexandrov2003,Galperin2005} A subsequent numerical analysis {\color{black} of systems under nonequilibirum conditions (with a finite voltage bias applied across the dot)} has revealed some  signatures of this phenomenon. However, so far the numerics were not able to supply conclusive evidence about the lifetime of the system in different conformational states of the molecule.\cite{PhysRevB.86.081412,PhysRevB.88.045137} On the other hand, there are several arguments against a bistable behaviour of such systems at low energies.\cite{Mitra2004,PhysRevB.89.205129,Ferdi2} The purpose of this paper is to reconsider the problem, trying to settle the open issues outlined above {\color{black} for systems in equilibrium}. By doing this we have made a twofold progress. Firstly, we report  path integral Monte Carlo (PIMC) simulations for the coordinate distribution functions of the localized vibrational degree of freedom, which are especially convenient for measurements in future experimental realizations of the model with the help of ultracold gas systems.
Furthermore, we show a natural extension of the adiabatic approximation, which is valid in the low energy sector. There, surprisingly, the system undergoes a crossover into a regime
which closely resembles the low energy limit of the Kondo model and 
which is determined by a single energy scale.

\begin{figure}
 \centering
 \includegraphics[width=.30\textwidth]{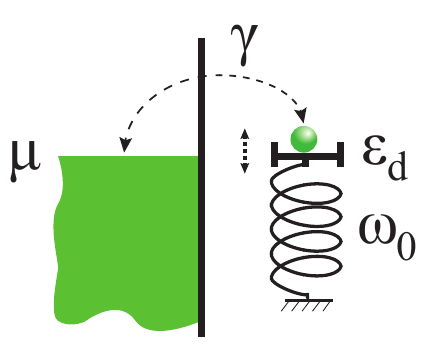}
 \caption{(Color online) \label{fig:setup} Schematic representation of the system described by the Hamiltonian \eqref{model}. 
}
 \end{figure}

The model for 
the molecular quantum dot 
possesses 
a single electronic level with energy $\epsilon_{\text{d}}$ and one vibrational mode, see Fig.~\ref{fig:setup} (we consider a spinless system and use units in which $\hbar=e=1$). The corresponding local phonon is just a harmonic oscillator with mass $m$ and frequency $\omega_0$. It is linearly coupled with strength $\lambda$ to the molecular electronic state which, in turn, is coupled via a tunnelling amplitude $\gamma^{}$ to the fermionic continuum in a metallic electrode kept at chemical potential $\mu=0$ at inverse temperature $\beta$. The Hamiltonian comprising the aforementioned features reads
 \begin{align}        \label{model}
    & H = H_\text{el}+H_\text{ph}+H_\text{int}\,,\\\nonumber
    & H_\text{el} = \epsilon_\text{d} \, d^\dagger d+\sum_k \left[\epsilon_k c^{\dagger}_k c_k +\left(\gamma \, c^{\dagger}_{k} d+\text{H.c.}\right)\right]\,,\\\nonumber
    & H_\text{ph} = \frac{P^2}{2m}+\frac{1}{2}m\omega_0^2 \, Q^2\,, \, \, \, \, \, 
     H_\text{int} = \lambda \, Q \, (d^\dagger d -1/2)\,.
   \end{align}
 $c_k$ and $d$ are the fermionic annihilation operators corresponding to the electronic states of the metallic electrode and of the molecule, respectively. $P$ and $Q$ are the momentum and position operator of the phonon.

Our main goal is to derive a numerical procedure to access the probability distribution of the expectation value of the phonon position operator for arbitrary parameters, e.~g. various coupling strength $\lambda$ and temperatures $T$, to search for signatures of bistability. In order to achieve this goal we first produce an effective action for the phonon by integrating out the electrode degrees of freedom out of \eqref{model} {\color{black} using standard functional integration technique}. The resulting action is given by
\begin{eqnarray}    \label{Slong}
 S &=& S_0[Q] + \int_0^\beta 
 d \tau d \tau' d^\dag(\tau) D_0^{-1}(\tau-\tau') d(\tau') 
 \nonumber \\
 &-& \lambda \int_0^\beta d \tau \, Q(\tau) \, d^\dag(\tau) d(\tau) \, ,
\end{eqnarray}
where $S_0[Q]$ is the action for a free harmonic oscillator and where by abuse of notation $d^\dag(\tau)$ and $d(\tau)$ denote adjoint variables. $D_0(\tau)$ is the Matsubara Green's function (GF) of the dot electron level, which can be found by elementary means. We concentrate on the wide flat band model, in which the conductance band of the electrodes is infinitely wide and has a constant density of states $\rho_0$.\footnote{This is not restrictive in any way. Any other band structure can be treated along the same lines.} Then one obtains
\begin{eqnarray}
D_0^{-1}(\Omega_n) = i\Omega_n - \epsilon_{\text{d}} - i \Gamma \, \mbox{sgn} \, \Omega_n
\end{eqnarray}
in energy representation with $\Omega_n = \pi (2n+1)/\beta$ and $\Gamma=\pi \rho_0 |\gamma|^2$ being the inverse lifetime of the electron on the dot. Further integrating out the dot's electronic degree of freedom yields for the partition function:
\begin{eqnarray}     \label{Zlong} 
  Z = \int {\cal D}Q \, e^{- S_0[Q] -
  \frac{1}{2} \ln \text{det} \left[ \beta \delta_{n m} D^{-1}(\Omega_n) + \lambda Q(\Omega_n - \Omega_m) \right]} \, 
\end{eqnarray}
 \begin{figure}
 \includegraphics[width=.47\textwidth]{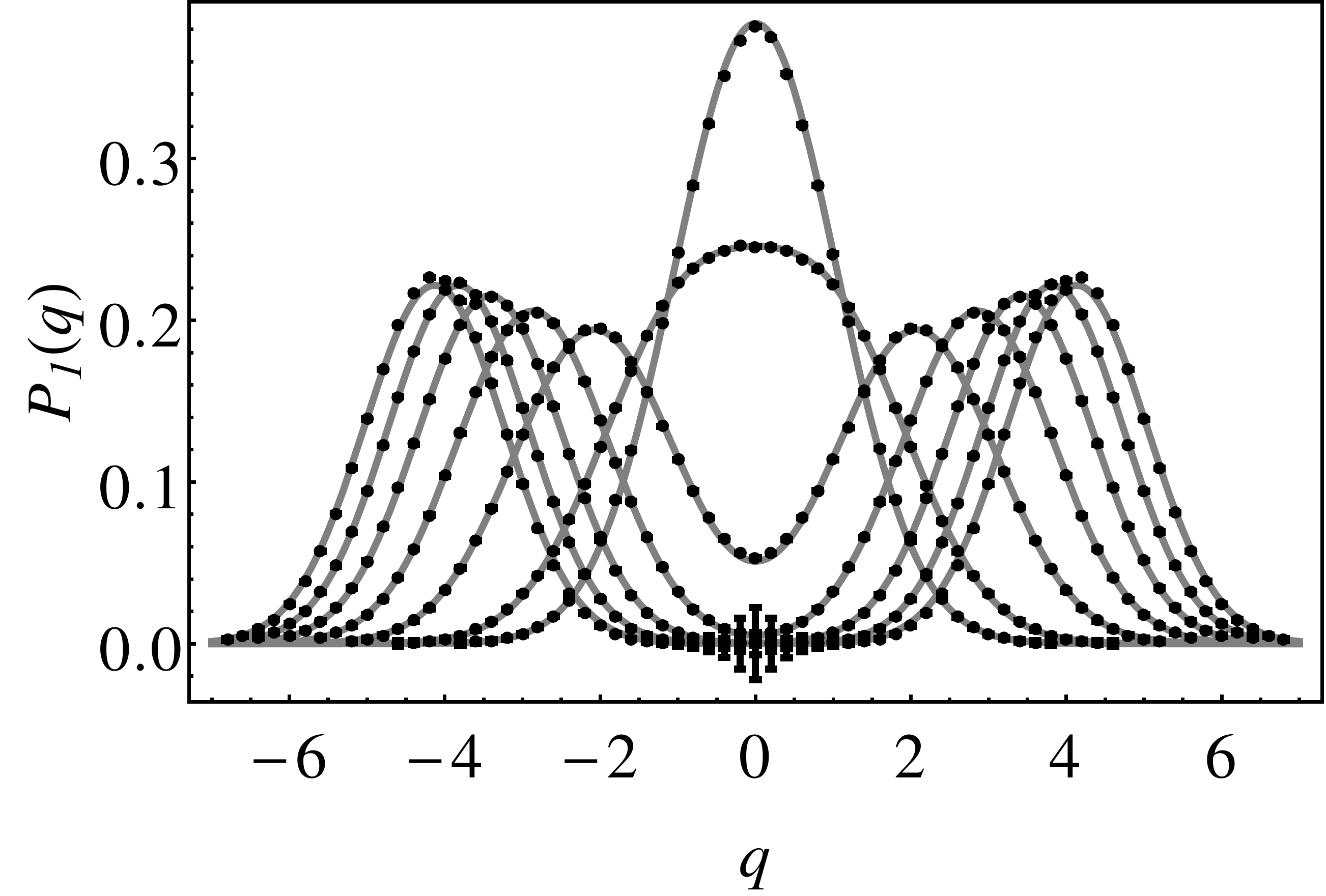}%
 \caption{Dependence of the phonon displacement probability distribution on
   electron-phonon coupling strength $\lambda$ obtained via PIMC simulations
   for $\beta\Gamma=\omega_0/\Gamma=1$. $\lambda$ increases from zero (curve
   with the highest maximum) to $\lambda=6$ (curve with the outmost maxima) in
   unit steps. 
   In all simulations $m \omega_0^2 =1$ so that $q_0$ and $\lambda$ are
   measured in units of $\sqrt{\Gamma/m \omega_0^2}$ and $\sqrt{\Gamma m
     \omega_0^2}$, respectively. 
   Dots represent the numerical data and curves are fits of a symmetric and
   normalized superposition of two Gaussians. In all simulations $\epsilon_{\text{d}}=0$.
}
\label{fig:lambda}
 \end{figure}

 This expression cannot be evaluated analytically. It is, however,
 amenable to numerical treatment and is the basis for our PIMC simulations. We
 follow the procedure outlined in
 [\onlinecite{PhysRevLett.100.176403,Werner2009,RevModPhys.83.349,MasterarbeitJuliane}]
 and adapt it for measurement of the irreducible cumulants of the type
 $q=\langle Q \rangle$ and $q^{(n)}=\langle (Q - \langle Q \rangle)^n \rangle$
 for $n>1$.  We use the notation $P_n(q^{(n)})$ for the corresponding distributions, which are the principal quantities of interest.

 The notation $\int {\cal D}Q$ within the imaginary-time path integral (\ref{Zlong}) symbolically refers to the continuum limit 
 \begin{eqnarray}\label{discrete}
  \lim\limits_{K\rightarrow\infty}\frac{1}{\cal N}\int \prod_{k=1}^{K-1} dQ_k \quad,
 \end{eqnarray}
where $K$ is the number of steps the imaginary time span of interest was devided into and $\Delta \tau$ is the corresponding step width. $Q_k$ indicates the phonon position at step $k$. 
${\cal N}$ denotes a normalization factor. It drops out of any observable since they all are given by ratios of limits of the type $\eqref{discrete}$.
For numerical purposes, the aforementioned limit is not performed, leaving the discrete version of (\ref{Zlong}) to be a high yet finite dimensional integral which may readily be evaluated by the means of Monte Carlo methods. The numerically evaluated mean of the phonon displacement $Q$ to some power $n$ for example reads
\begin{align}\label{Qn}
 \left\langle Q^n\right\rangle&=\frac{1}{\cal N}\int \prod_{k=1}^{K} dQ_k\,Q_K^n\,e^{-\frac{m \Delta\tau }{2}\sum\limits_{k=1}^{K}\left[\frac{(Q_k-Q_{k-1})^2}{(\Delta \tau)^2}+\omega_0^2Q_k^2\right]}\,\\\nonumber
&\times\,\det\left[\left(\begin{array}{c}\delta_{n m}- \Delta \tau\lambda\,Q_k\,G_{0,n-m}^{\mathcal{T}} \end{array}\right)\right]\quad,
\end{align}
where $G_{0,n-m}^{\mathcal{T}}$ is the Fourier transform of the thermal analogon $G_0^{\mathcal{T}}(\tau)$  of the time-ordered GF. The latter is connected to the Matsubara GF $D_0(\tau)$ by Keldysh rotation, resulting for our model in
\begin{align}\nonumber
G_0^{\mathcal{T}}(\tau)&=\theta(\tau)G_0^{>}(\tau)+\theta(-\tau)G_0^{<}(\tau)\quad,\\\label{th_G}
G_0^{>}(\tau)&=\int_{-\infty}^\infty \frac{d\omega}{\sqrt{2\pi}} \frac{2\Gamma e^{-\omega\tau}}{(\omega-\epsilon_\text{d})^2+\Gamma^2}\left[f(\omega)-1\right]\quad,\\\nonumber
G_0^{<}(\tau)&=\int_{-\infty}^\infty \frac{d\omega}{\sqrt{2\pi}} \frac{2\Gamma e^{-\omega\tau}}{(\omega-\epsilon_\text{d})^2+\Gamma^2} f(\omega)\quad .
\end{align}
Here, $f(\omega)$ denotes the Fermi distribution function of the electrode. Note that $G_0^{<}(\tau)$ is only finite for $-\beta<\tau<0$ and $G_0^{>}(\tau)$ for $0<\tau<\beta$. The Heaviside step function must be $\theta(0)=1/2$ at the origin.

Applying Monte Carlo method to (\ref{Qn}) in order to estimate it simply means stochastically sampling its integrand -- i.~e. randomly travelling phase space and evaluating the integrand at the points visited. If samples were drawn uniformly (which for an infinite interval is rather a challenge), numerical effort would scale exponentially with the dimension of the space one wishes to sample.  We employ an importance sampling procedure instead, which uses the fact that the
physical distribution functions $P_n\left(q^{(n)}\right)$ actually have a rather small support -- interactions and Boltzmann factors reduce the relevant phase space volume significantly. Thus samples may not be drawn uniformly but according to a distribution $P_\text{MC}(\{Q_k\})$, such that only the dominant areas of the phase space are sampled with high accuracy. The non-uniform exploration of phase space must be countered by accumulating the ratio of the integrand and the probability density it is drawn from instead of the plain integrand.
This strategy is applied optimally if $P_\text{MC}(\{Q_k\})$ 
is identical 
to the integrand of the integral to be estimated. In our case this optimum can be reached since due to the thermal type of our problem the integrand is non-negative and normalizable.
However, this cannot be done in case of a system in non-equilibrium, when, for instance, the quantum dot is coupled to two metallic electrodes with different chemical potentials.
 In order to produce samples distributed according to the integrand of (\ref{Qn}) we employ the Metropolis-Hastings algorithm.\cite{Metropolis1953,Hastings1970} That is, we construct a Markov chain of paths $\{Q_k\}$ whose equilibrium distribution is equal to the one we are looking for. Thus, once equilibrated, the Markov walk produces a manifold of trajectories, since any step of it represents an entire phonon path.

 Both the systematic error due to the finite number of discretization steps and the statistical error due to stochastic nature of simulations can be made arbitrarily small by  increasing the discretization resolution and number of samples, respectively.
%
%
{\color{black} In addition, Markov chain methods are a source of yet another error
type. The obtained random samples are not statistically independent
since they are not drawn from a distribution but constructed from the
Markov walk. This walk paces configuration space, rendering its
consecutive steps correlated.
}
That implies a twofold difficulty: first of all, variances are underestimated if calculated in the same way as for independent numbers and, secondly, there is  no means of assessing the convergence of individual walks. 
Only if two points are separated by more then the autocorrelation time, they can be considered being statistically independent. 
 In order to control that we use the procedure
 proposed by Flyvbjerg, \cite{Flyvbjerg1989} according to which samples are paired, pair averages are treated as independent samples and the statistical error is estimated. Subsequently, the just obtained averages are paired once again and the error is estimated anew, and so forth. The iteration is stopped, once the estimators form a plateau.  In this way we obtain very reliable numerical data, the overall error of which is extremely small.

 Fig. \ref{fig:lambda} shows the distribution of the average value $\langle Q
 \rangle$ for increasing electron-phonon coupling. For $\lambda=0$ we deal with
 a simple harmonic oscillator and $P_1(q)$ is given by
 the probability distribution of the groundstate wave function. For growing
 interaction strength one observes a splitting of the central maximum into two
 resulting in a bimodal distribution. 
  This can be interpreted as a formation of a double dip in the oscillator's effective potential, the latter in close analogy to the uncoupled case, where one would expect $P_1(q) \sim
\exp[-\beta V(q)]$ in the semiclassical limit, with $V(q)$ denoting the oscillator's potential energy.
 This is closely related to the bistability effect
 discussed earlier.\cite{my,Galperin2005} An approximation used in these works
 assumes the phonon degrees of freedom to be much slower than the electronic
 ones, e. g. when the typical timescale for the electron dynamics on the dot
 $\sim \Gamma^{-1}$ is much smaller than the phonon oscillation period $
\omega_0^{-1} \gg \Gamma^{-1}$ one can use the adiabatic, or Born-Oppenheimer
 approximation. Then the field $Q(\tau)$ in the $\lambda$-term in \eqref{Zlong}
 can be taken to be static: $Q(\Omega_n - \Omega_m)= \delta_{n m} \beta Q$ with
 constant $Q$. The resulting functional integral can then be evaluated by virtue of the saddle point approximation, which 
is performed in [\onlinecite{my}], yielding in the
 case of strong coupling $\lambda^2 \gg m \omega_0^2 \Gamma$ for 
$P_1(q)$ two $\delta$-shaped maxima at the positions
$Q_\pm = \pm q_{\mbox{max}}$ where $q_{\mbox{max}} = \lambda/2 m \omega_0^2$. 
It turns out that a careful evaluation of the adiabatic approximation beyond
the saddle point is indeed able to recover the actual shape of the distribution
function obtained numerically with high degree of accuracy. Moreover, the
prediction for $Q_\pm$
is perfectly reproduced by our simulations. {\color{black} In general the central dip in $P_1(q)$ at the original equilibrium position of the uncoupled oscillator $q=0$ might or might not touch the axis of ordinates. We call the former case \emph{perfect bistability} while in the latter situation we are dealing with the \emph{spurious bistability}.}

\begin{figure}
 \includegraphics[width=.475\textwidth]{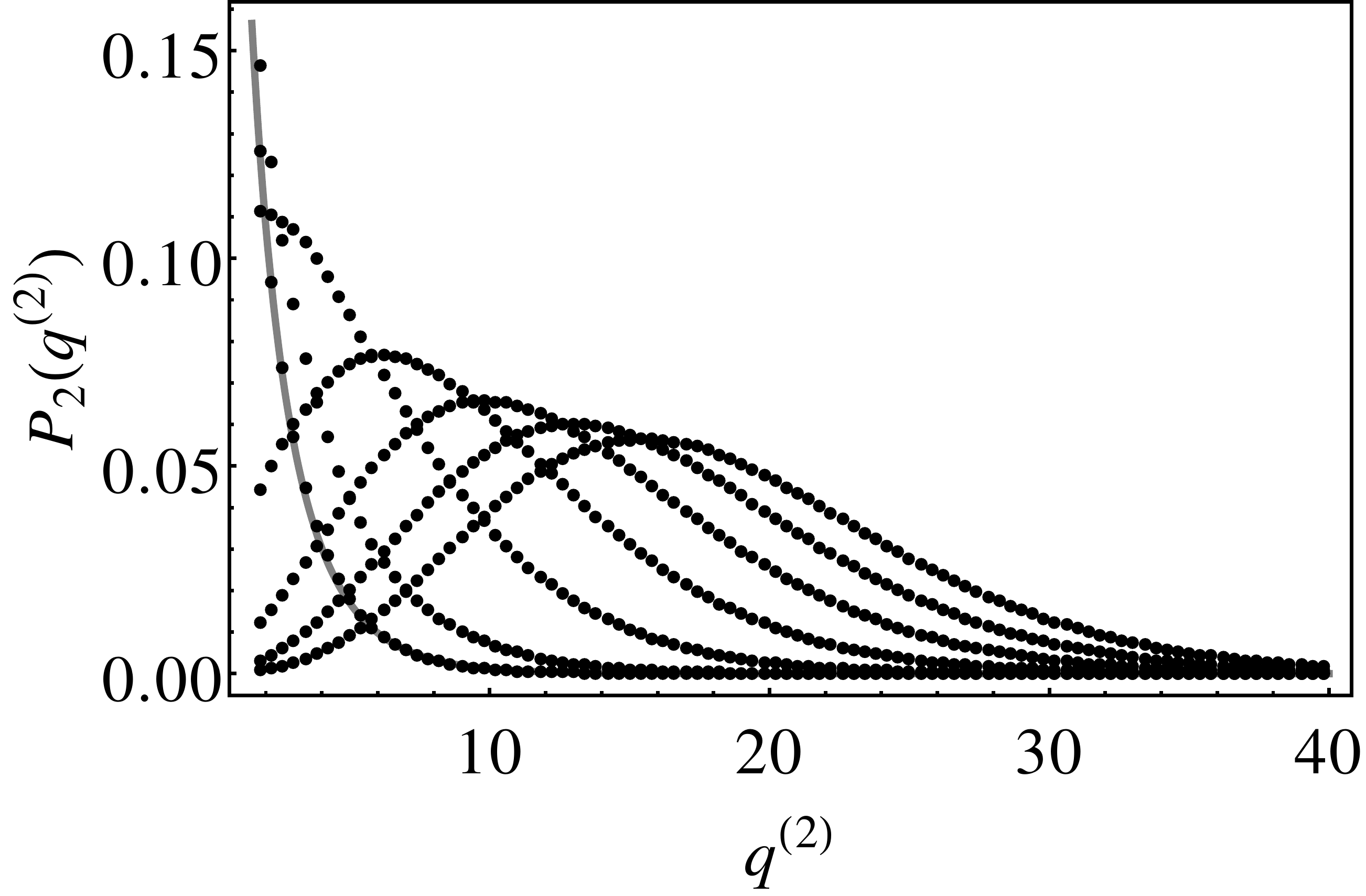}%
 \caption{
 Distribution function of the variance $q^{(2)}$ for different electron-phonon strengths: $\lambda=0$ (the corresponding data points perfectly coincide with the prediction of \eqref{Pn}, plotted as a curve), and $\lambda=1,2,3,4,5,6$ (the maxima of the respective data sets lie at larger $q^{(2)}$ for growing $\lambda$).
 }\label{fig:P_2}
 \end{figure}

While in the uncoupled case the distribution $P_n(q^{(n)})$ of higher moments for $n>1$ can trivially be found from $P_1(q)$ of $q=\langle Q \rangle$ via the relation 
\begin{eqnarray}   \label{Pn}
 P_n(q^{(n)}) = P_1(q)/(n q^{n-1}) \, ,
\end{eqnarray}   
it is not possible any more in the case $\lambda \neq 0$ because in a fully interacting system a multitude of irreducible multi-particle correlations emerge and the single-particle picture, which is essential for the derivation of \eqref{Pn} breaks down.  This is demonstrated in Fig.~\ref{fig:P_2}, where we have plotted the distribution $P_2(q^{(2)})$ of the second cumulant for an interacting system as well as the prediction \eqref{Pn}.

\begin{figure}
\centering
 \includegraphics[width=.45\textwidth]{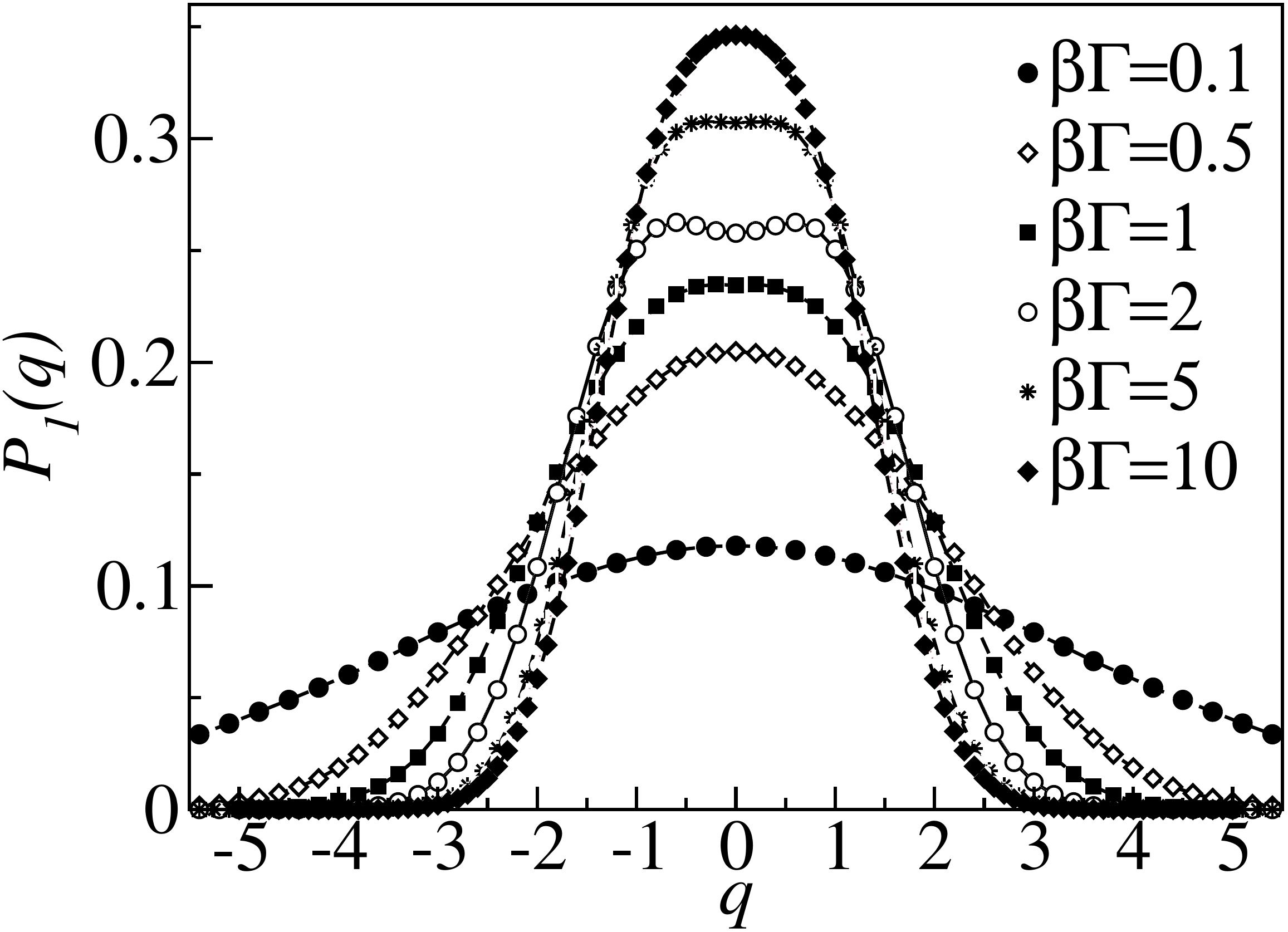}
\caption{
Phonon displacement probability distribution for $\lambda=1$ and $\omega_0/\Gamma=1$ for different inverse temperatures $\beta$. \label{FigLambda1}}
\end{figure}
\begin{figure}
\centering
 \includegraphics[width=.45\textwidth]{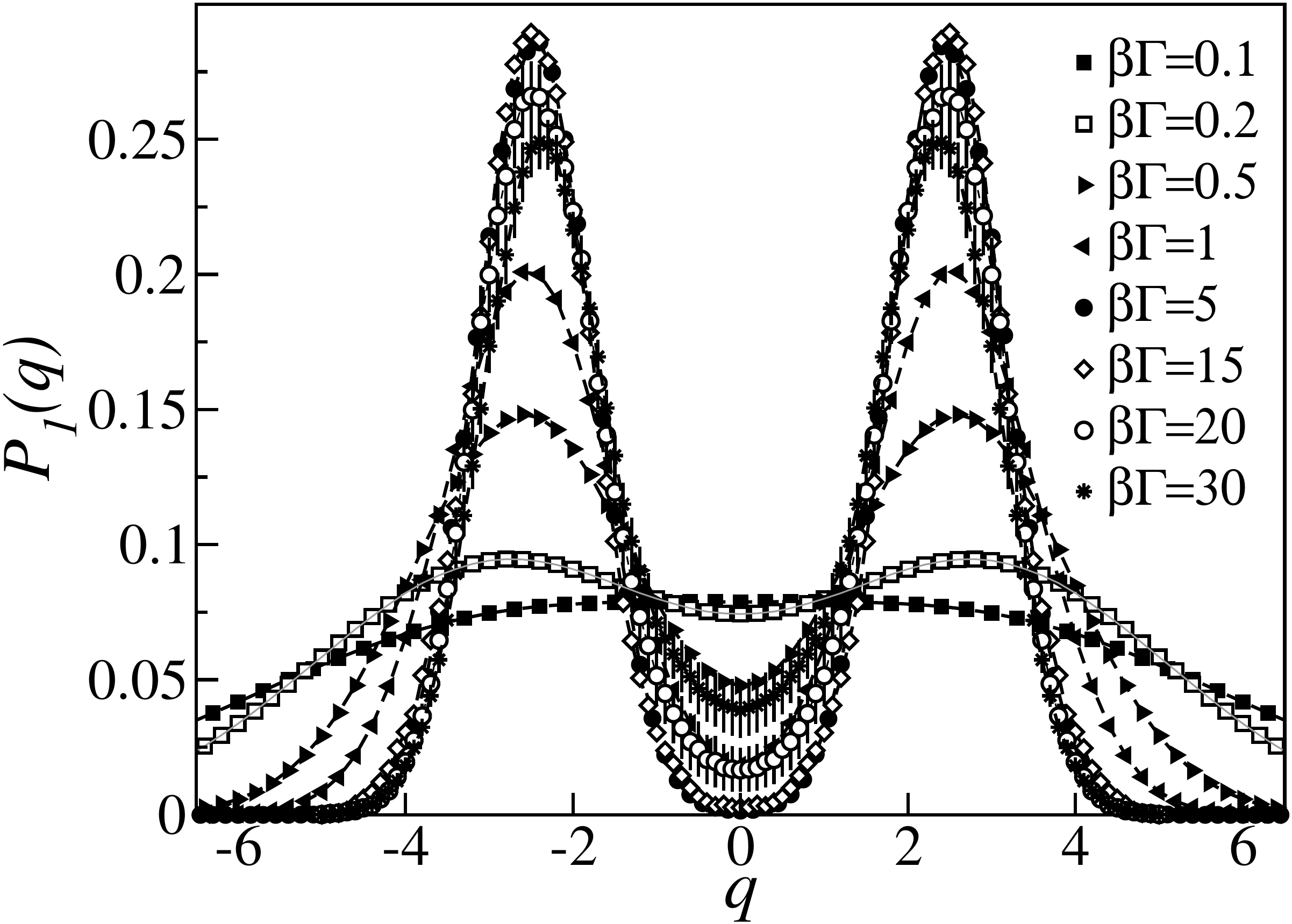}%
\caption{
Phonon displacement probability distribution for $\lambda=2.5$ and $\omega_0/\Gamma=1$ for different inverse temperatures $\beta$. 
\label{FigLambda25}}
\end{figure}

Upon further lowering the temperature one observes a rather fast collapse of
the bimodal distribution function to one with a single maximum at $Q=0$.
This kind of crossover
occurs at a very low critical temperature, which strongly depends on $\lambda$
and can be seen in PIMC results for two different values of the coupling,
$\lambda=1$ and $2.5$. Fig.~\ref{FigLambda1} shows the data for intermediate
coupling strength $\lambda=1$. 
Here the distribution at $\beta \Gamma=2$ is
bimodal due to the bistability effect, while $P_1(q)$ exhibits one unique maximum
for all investigated temperatures
above or below $\beta\Gamma=2$.
The collapse of the {\color{black} spurious} bistability at low
temperature can be even more clearly observed in a strongly coupled system with
$\lambda=2.5$, see Fig.~\ref{FigLambda25}. It turns out that under these
conditions the bistability starts to emerge already at $\beta\Gamma \approx
0.2$ and becomes almost perfect at $\beta\Gamma \approx 2$ (in a sense that then $P(0)\approx 0$,  which means that the probability
  to find the phonon localized in the original
  equilibrium position of the uncoupled oscillator vanishes completely), persisting this way all the way down to
temperatures $\beta \Gamma \approx 15$. After that the bistabilty starts to
degrade as demonstrated by the curves for $\beta\Gamma \approx
  20$ and $\beta\Gamma \approx
  30$ , which no longer touch the ordinate axis at $Q=0$.

This interesting behaviour leaves open a number of important questions. The
first and most important of them regards the dependence of the lower crossover
temperature $T_c$,  below which $P_1(q)$ again
  shows one unique maximum, on the electron-phonon coupling strength.
The adiabatic approximation is not able to estimate $T_c$.
In what follows we improve it
  by taking care of the low-energy fluctuations of the $Q(\tau)$ field
 in the
functional integral \eqref{Zlong}. In order to make progress it is helpful to
return to the action \eqref{Slong}. Here $Q(\tau)$ plays a role of a time-dependent
potential for the local fermion. In that respect the problem in question
formally resembles the X-ray edge problem solved by Nozi{\`{e}}res and De
Dominicis (ND), with our local fermion being the equivalent to
the core-hole electron level of the ND problem.\cite{PhysRev.178.1097} This
similarity has been recognized in the seminal paper by Yu and Anderson
(YA),\cite{PhysRevB.29.6165} in which the authors considered the scaling
behaviour of a fermionic continuum locally coupled to an Einstein phonon, and
in which they used Hamann's version of the ND solution.\cite{PhysRevB.2.1373}
Although our model is not fully equivalent to those treated in
any of the works mentioned above, we can easily adapt their mathematical
apparatus to our needs. Despite the fact that the resulting effective model is equivalent to the spin-boson Hamiltonian, \cite{Leggett} its derivation is much simpler using the approach inspired by YA.

\begin{figure}
 \centering
 \includegraphics[width=.4\textwidth]{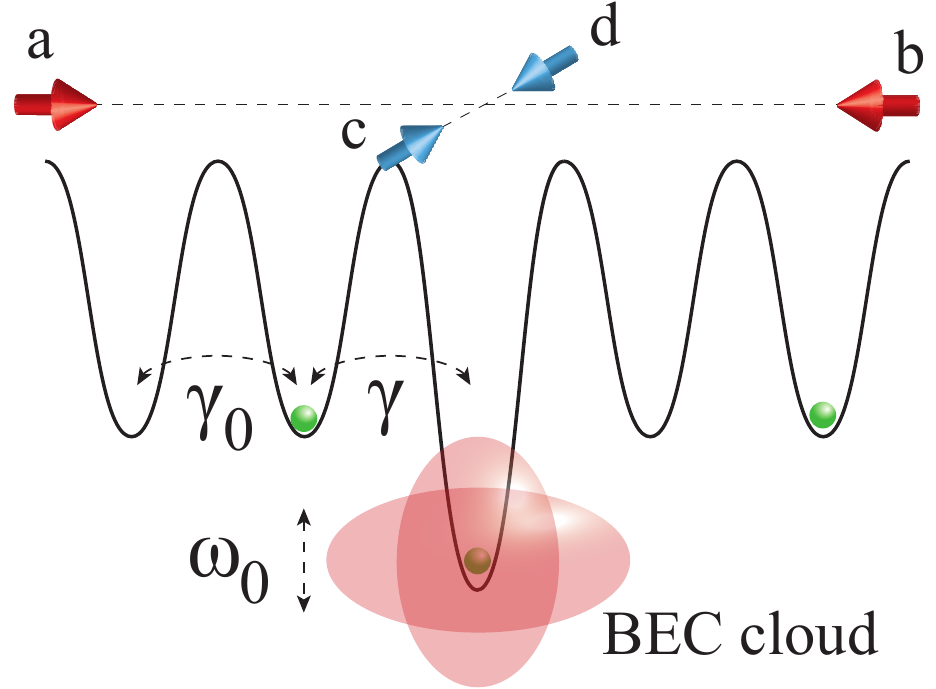}
 \caption{(Color online) \label{fig:BECsetup} Synthetic quantum system on the basis of an ultracold boson-fermion mixture, which is described by \eqref{model}. Two laser fields arranged along the axes $ab$ and $cd$ create a deep regular lattice for the fermions, which plays the role of the continuum (the electrode in the molecular quantum dot realization). Tunnelling between the lattice sites is equal and given by $\gamma_0$.  By an auxiliary laser field an impurity site is produced, the tunnelling amplitude $\gamma$ to which is different from $\gamma_0$. The impurity site is immersed in a BEC cloud, with which it can interact and thereby excite lowest lying eigenmodes of the cloud.}
 \end{figure}

The idea is based on integrating out the local fermion field, which is done in imaginary time rather than in the energy domain as in \eqref{Zlong}. To that end we rewrite the full system action as {\color{black} (see also Refs. [\onlinecite{PhysRevB.29.6165}],[\onlinecite{PhysRevB.2.1373}])}
\begin{eqnarray}  \nonumber
 e^{-S(Q)} =  e^{-S_0(Q)} \langle e^{-\lambda \int_0^\beta d \tau \, Q(\tau) d^\dag(\tau) d(\tau)}
\rangle
  = e^{-S_0(Q)} Z_\psi \, , 
\end{eqnarray}
where the average is taken over the electronic degrees of freedom assuming the
phonon path $Q(\tau)$ being fixed. The computation of $Z_\psi$ is accomplished
in the usual way by differentiating it with respect to the coupling constant
$\lambda$, performing the average and integrating again with
  respect to $\lambda$.{\cite{Abrikosov1975,PhysRevB.2.1373} As a result one obtains
\begin{eqnarray}
  Z_\psi = \exp\left[
-\int_0^\lambda d \lambda \int_0^\beta d \tau Q(\tau) D(\tau, \tau+0^+)
 \right] \, ,
\end{eqnarray}
where 
 $D(\tau,\tau') = - \langle T_\tau d(\tau) d^\dag(\tau') \rangle_\lambda$
 is the Matsubara GF for the local fermion, calculated \emph{in presence} of
 the potential $Q(\tau)$. 
 It is best computed as a
 solution of the following exact Dyson equation:
\begin{eqnarray}   \label{Dyson}
 D(\tau,\tau') &=& D_0(\tau-\tau') + \lambda \int_0^\beta d \tau'' \, D_0(\tau-\tau'') 
\nonumber \\
&\times& 
Q(\tau'')  D(\tau'',\tau')
 \, .
\end{eqnarray}
From now on we start making approximations. As was realized by ND, in order to
access the low-energy behaviour of our system it is sufficient to get hold of
the long-time asymptotics of the solution of Eq.~\eqref{Dyson}.\cite{PhysRev.178.1097} This can
be conveniently done with the help of the regularized version of
\begin{eqnarray}
D_0(\tau)
\approx - \frac{\Gamma}{\epsilon_d^2 + \Gamma^2} \left[ \frac{\cal P}{\pi \tau} +
 \frac{\epsilon_d}{\Gamma} \, \delta(\tau) \right] \, , 
\end{eqnarray}  
where ${\cal P}$ denotes the
principal value.  With this simplification the solution of the Dyson equation
can be taken from [\onlinecite{19703}]. As a result, using
the notation $\xi(\tau) = \lambda Q(\tau)/\Gamma$ we obtain 
\begin{eqnarray}
Z_\psi =
\exp\left\{-\int_0^\beta d \tau \, [V_1(\xi) + T_1(\xi)] \right\} \, , 
\end{eqnarray}
where
\begin{eqnarray}
V_1(\xi) = - (2 \Gamma/\pi)[\xi \arctan \xi - (1/2) \ln (1 + \xi^2)]
\end{eqnarray}
is an addition to the original harmonic oscillation potential, giving rise to precisely the same terms, which are responsible to the adiabatic (Born-Oppenheimer) approximation.\cite{my} The other, \emph{transient}, term describes retardation effects and represents the next order expansion around the adiabatic approximation,\cite{PhysRevB.2.1373}
\begin{eqnarray}
 T_1(\xi) &=& \frac{\cal P}{\pi^2} \int_0^\beta d \tau' \, \frac{1}{\tau - \tau'} \, \xi(\tau) \, \frac{d \xi(\tau')}{d \tau'} \,
 \nonumber \\ &\times&
  \frac{1}{\xi^2(\tau) - \xi^2(\tau')} \, \ln \left[
 \frac{1 + \xi^2(\tau)}{1 + \xi^2(\tau')}
 \right] \, .
\end{eqnarray}
The partition function of our system is now a functional integral over
$\xi(\tau)$ and the oscillator momentum variable $P(\tau)$. It was
recognized by Hamann, that in the regime, in which the
oscillator potential $m\omega_0^2 Q^2(\tau)/2 + V_1(Q(\tau))$ develops two
distinct minima (bistability regime) the dominant paths are given
  by hopping events between them.\cite{PhysRevB.2.1373} It was later shown by
YA that this picture is only insignificantly altered by the kinetic term of the
oscillator.\cite{PhysRevB.29.6165} 
  Following YA and
  modelling the hopping events in the same way one can explicitly write down
the partition function for the system in question. It turns out to be identical
to the result in Eq.~(61) of YA in which the hopping fugacity is replaced by
\begin{eqnarray}
 y &=& \frac{2^{-\alpha/2}}{\xi_0} \, \exp\left[-2 m \left(\frac{\Gamma}{\lambda} \right)^2 \frac{\xi_0^2}{\tau_0} \right] \,
 \\  \nonumber
&\times& \exp\left\{-\tau_0 \left[ \frac{m \omega_0^2}{3}\left(\frac{\Gamma}{\lambda} \right)^2 \xi_0^2 - \frac{\Gamma}{\pi} (1 - \frac{1}{\xi_0} \arctan \xi_0) \right]\right\} \, ,
\end{eqnarray}
where $ \alpha = \left[ (2/\pi) \arctan \xi_0 \right]^2$, $\xi_0 = \lambda
q_{\mbox{max}}/\Gamma$ and $\tau_0$ is the duration of the hop, which is found
from precisely the same prescription as the one used by YA. In the strong
coupling limit the partition function for the system formally equals the one
found in [\onlinecite{PhysRevB.1.4464}] for the Kondo problem. It 
 possesses a non-trivial scaling behaviour towards low energies, which is
reflected in a surge of the number of hopping events. Below the
Kondo temperature $T_K$ the system undergoes a transition to a new kind of
ground state, 
  which is
universally characterized by $T_K$. From the equivalence of the partition
functions we find the following relation {\color{black} for the temperature below which the bistability vanishes}:
\begin{eqnarray}
   \label{TK}
 T_c \approx C (\lambda^2/m \omega_0^2) e^{-\pi \lambda^2/(8 \Gamma m \omega_0^2)} \, ,
\end{eqnarray}
where $C$ is some yet unknown numerical prefactor. 
When adopting the hybridization $\Gamma$ as our energy unit $T_c$ becomes a \emph{single parametric function} of $x=\lambda^2/(\Gamma m \omega_0^2)$: 
 $\beta_c \Gamma \approx (1/C x) e^{\pi x/8}$.
From the comparison with the data presented in Fig.~\ref{FigLambda25} we conclude that $\beta_c \Gamma > 30$ and therefore we can estimate the constant $C$ as being smaller than $0.06$.

Bistability signatures have recently been observed in nonequilibrium electron transport through molecular quantum dots. Although the corresponding simulations have been carried out at zero temperature no effects related to Kondo physics have been seen. 
{\color{black} It can be attributed to the finite bias voltage playing the role of effective temperature.\cite{Urban}  On the other hand, there is also another reason for absence of Kondo signatures in case of small voltages.}
Starting with an initially decoupled system, which is the case in the mentioned nonequlibrium simulations, it takes a finite time $t_K \sim 1/T_K$ for the Kondo effect to become fully established.\cite{PhysRevLett.83.808}  An estimation shows that $t_c \sim 1/T_c$ is of the same order or even larger than timescales addressed by the simulations. Thus the genuine stationary state has not yet been achieved, which would explain why the bistability signatures were visible. In [\onlinecite{FerdiDiss}] the authors reported a bistability collapse on a large time scale, which is consistent with our estimation.  

 We would like to mention, that although both the numerics as well as the analytical discussion are performed for the resonant case $\epsilon_{\text{d}}=0$, we expect everything to hold also for finite not too large values of $\epsilon_{\text{d}}$. This assertion definitely can be shown to hold for the bistability effect by a direct computation.\cite{my} As far as the Kondo crossover is concerned, it is known to survive in a finite magnetic field, which is smaller than $T_K$. From the mapping between the models one finds that $\epsilon_{\text{d}}$ plays the role of the magnetic field. Therefore we conclude that for $\epsilon_{\text{d}} < T_c$ one should expect the Kondo crossover to be seen as well.

Systems of primary interest, in which the bistability and YA-type Kondo effect should be observable and preferrably also find practical applications are contacted molecules. Here the fundamental obstactle is the weak electron-phonon coupling.
 However, this restriction is less severe in systems, which are based on carbon nanotubes. As reported in [\onlinecite{Leturcq:2009fp}] the dimensionless electron-phonon coupling strength in such systems can exceed $\sim 3$, which is sufficient for the bistability as well as Kondo crossover to become observable. The only reason why that has not yet been observed experimentally is the insufficiently low temperature. This difficulty can certainly be overcome in future experiements.

Another class of systems, in which strong electron-phonon coupling is realizable, are ultracold gas mixtures. The authors of Ref.~[\onlinecite{Zwierlein_polaron}] have succeded in producing a synthetic quantum system with an effective electron-phonon coupling strength, which would be sufficient to observe the phenomena we have discussed above. In order to see them} we envision an optical lattice which can be loaded with fermionic atoms, e. g. with $^{6}$Li, which plays the role of a fermionic continuum, see Fig.~\ref{fig:BECsetup}. 
Additionally a trap for a Bose-Einstein condensate (BEC), e.~g. of $^{23}$Na atoms, is positioned in the vicinity of one of the lattice sites,
such that there is at least a partial spatial overlap between the BEC in its groundstate and one (or several) of the fermionic sites (a prototype system is realized in [\onlinecite{PhysRevLett.111.070401}]).  The latter then plays the role of the dot energy level. In presence of boson-fermion interactions there would be a coupling $\lambda \neq 0$ between the lowest-lying harmonic mode of the BEC and the localized fermion level, such that the model \eqref{model} applies. Due to the high tunability of such systems we expect that not only the bistability regime might be reached, but also the much deeper lying Kondo fixed point. The simplest observable of interest is then the spatial dimension of the BEC droplet, which is proportional to $\langle Q \rangle$ and its cumulants we have discussed above. 

To conclude,
we have analyzed the low energy limit of the molecular quantum dot model in equilibrium. With the help of a path integral Monte Carlo method we have computed the probability distribution functions of different observables of the harmonic degree of freedom. We have shown, that in the strong coupling regime at not too low temperature the distribution of the expectation value of the average oscillator coordinate becomes bimodal. This is a consequence of the bistability effect reported previously, which can be described in the framework of the Born-Oppenheimer (adiabatic) approximation. Upon lowering the temperature the system crosses over into the true low energy fixed point, in which the bistability vanishes. By means of an analytical expansion around the adiabatic approximation we have shown that the emergent behaviour is connected to the Kondo effect, which is characterized by a single parameter -- the crossover temperature $T_c$ -- and have related it to the parameters of the system under investigation. 
Furthermore we have discussed the experimental implications of the predicted phenomena and suggested a setup for their observation on the basis of ultracold boson-fermion mixtures.

The authors would like to thank T. Novotn{\'{y}}, H. Grabert, and K. F. Albrecht for many intersting discussions. LM and AK aknowledge the support by the Deutsche Forschungsgemeinschaft (Germany) under Grants No. MU 2926/1-1 and No. KO 2235/5-1. LM further acknowledges computational support from the Black Forest Grid initiative (BFG).

\bibliography{YuAnderson_citing_bibliography}
\end{document}